\documentclass[12pt]{article}

\usepackage{amsmath}
\usepackage{graphicx,psfrag,epsf}
\usepackage{enumerate}
\usepackage{natbib}
\usepackage{url} 
\usepackage{mathrsfs,amsfonts,amssymb,amsthm}
\usepackage{mdwmath}
\usepackage{mdwtab}
\usepackage{color,soul}
\newcommand{\blind}{0}

\newtheorem{proposition}{Proposition}

\newcommand{\E}{{\bf E}}
\newcommand{\Prob}{{\bf P}}
\newcommand{\Cov}{\textbf{Cov}}

\def\spacingset#1{\renewcommand{\baselinestretch}%
{#1}\small\normalsize} \spacingset{1}

\newcommand{\X}{\textbf{X}}
\newcommand{\Y}{\textbf{Y}}
\newcommand{\bfmu}{\boldsymbol {\mu}}
\newcommand{\bfxi}{\boldsymbol {\xi}}
\newcommand{\bfeta}{\boldsymbol {\eta}}
\newcommand{\bftheta}{\boldsymbol {\theta}}
\newcommand{\bfepsilon}{\boldsymbol {\epsilon}}

\begin{document}


\if0\blind
{
  \title{\bf Thresholded Multivariate Principal Component Analysis
for Multi-channel Profile Monitoring}
  \author{ Yuan Wang,  Kamran Paynabar, Yajun Mei   
    \hspace{.2cm}\\
    \small H. Milton Stewart School of Industrial and Systems Engineering, Georgia Institute of Technology}
  \maketitle
} \fi

\if1\blind
{
  \bigskip
  \bigskip
  \bigskip
  \begin{center}
    {\LARGE\bf  Thresholded Multivariate Principal Component Analysis
for Multi-channel Profile Monitoring}
\end{center}
  \medskip
} \fi



\bigskip
\begin{abstract}
Monitoring multichannel profiles has important applications in  manufacturing systems improvement, but
it is non-trivial to develop efficient statistical methods due to two main challenges.
First, profiles are high-dimensional functional data with intrinsic inner- and inter-channel correlations,
and one needs to develop a dimension reduction method that can deal with such intricate correlations for the purpose of effective monitoring.
The second, and probably more fundamental, challenge is that the functional structure of multi-channel profiles might change over time, and thus
the dimension reduction method should  be able to automatically take into account the potential unknown change. 
To tackle these two challenges, we propose a novel thresholded multivariate principal component analysis (PCA) method for multi-channel profile monitoring. Our proposed method consists of two steps of dimension reduction: It first applies the functional PCA to extract a reasonable large number of features under the normal operational (in-control) state, and then use the soft-thresholding techniques to further select significant features capturing profile information in the out-of-control state. The choice of tuning parameter for soft-thresholding is provided based on asymptotic analysis, and extensive simulation studies are conducted to illustrate the efficacy of our proposed  thresholded PCA methodology.
\end{abstract}

\noindent%
\vfill

\newpage
\spacingset{1.45} 


\section{Introduction}\label{sec:intro}

Profile monitoring plays an important role in manufacturing systems
improvement (\cite{noorossana2011statistical}, \cite{qiu2013introduction}),
and a standard setup is to monitor a sequence of profiles (e.g.  curves or functions)
over time to check whether the underlying functional structure of
the profiles changes or not. Extensive research has been done for
monitoring {\it univariate} profile or {\it real-valued} functions
in the area of statistical process control (SPC) in the past decades,
and standard approaches are to reduce the univariate profiles in
the infinite-dimensional or high-dimensional functional spaces to a low-dimensional set of
features (e.g., shape, magnitude, frequency, regression coefficients,
etc.). See, for instance, work by \cite{jin2000diagnostic},  
\cite{ding2006phase},
\cite{jeong2006wavelet}, \cite{jensen2008monitoring}, \cite{berkes2009detecting}, \cite{chicken2009statistical},
\cite{qiu2010nonparametric}, 
\cite{abdel2013semiparametric}.

\begin{figure}
\centering 
 \includegraphics[width=3.5in]{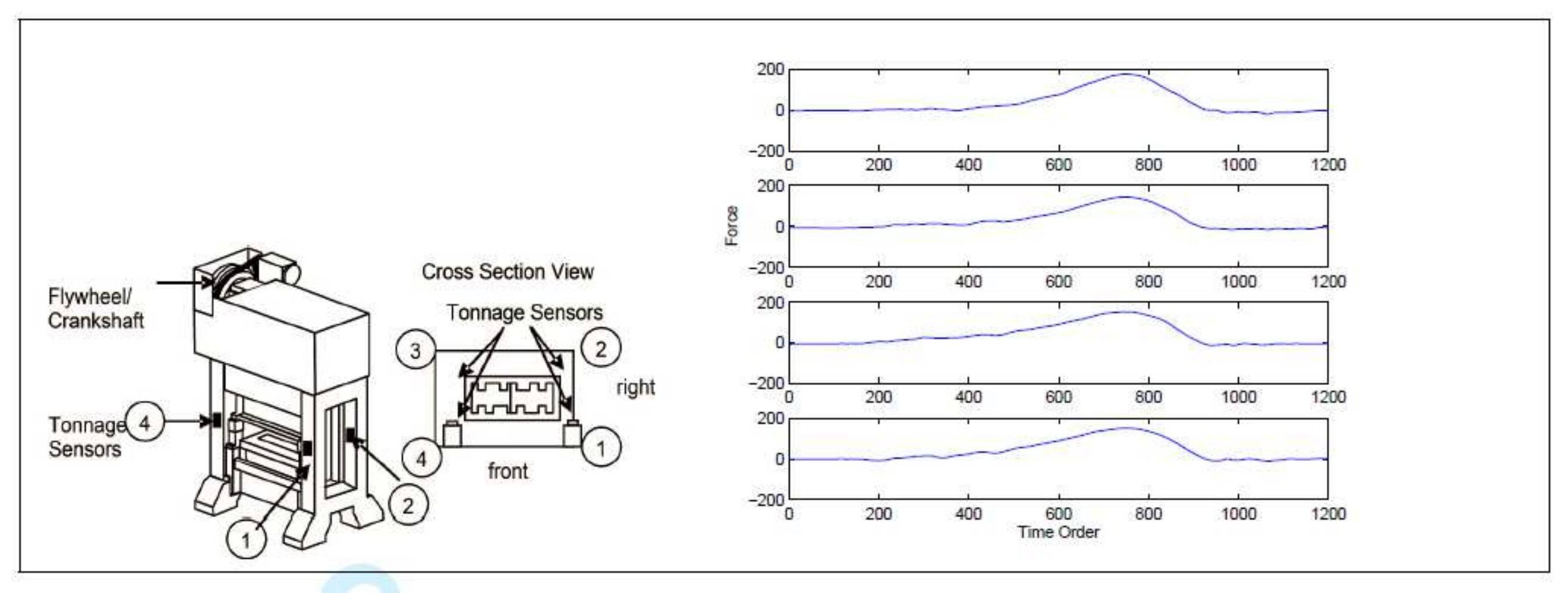}\\
 \protect\caption{\textit{: Left:} A forging machine with 4 tonnage sensors. \textit{Right:}
A single run sample of four-dimensional functional data.}

\label{fig:introa}
\includegraphics[width=3.5in]{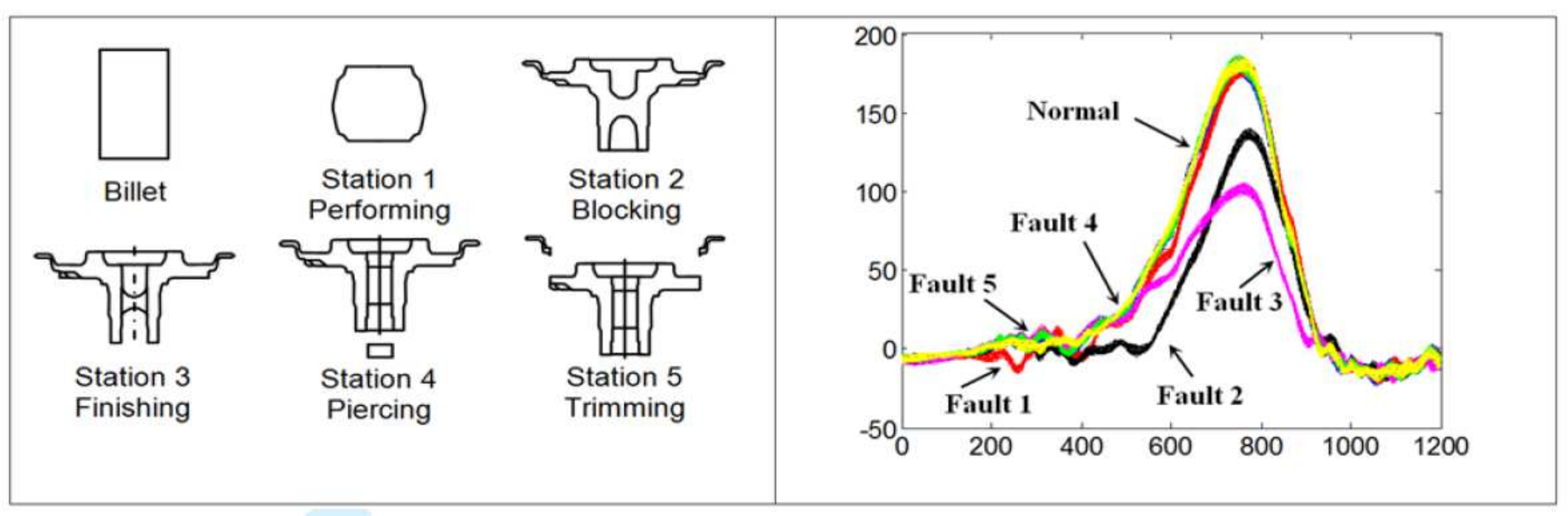}\\
 \protect\caption{\textit{: Left:} Shape of workpieces at each operation. \textit{Right:}
Tonnage profile for normal and missing operations.}

\label{fig:introb}

\end{figure}

Nowadays manufacturing systems are often equipped with a variety of
sensors capable of collecting several profile data simultaneously,
and thus one
often faces the problem of monitoring multichannel or multivariate
profiles
that have rich information about systems performance.  A concrete motivating example of this paper
is from a forging process, shown in Figure \ref{fig:introa} and \ref{fig:introb}, in which
multichannel load profiles measure exerted forces in each column of
the forging machine. Here each data is a four-dimensional vector
function or four curves that have similar but not identical
shapes when the machine is operating under the normal state.
While various methods have been developed for univariate profile monitoring,
they often cannot easily be extended to
multichannel profiles, and research on monitoring multivariate/mutichannel
nonlinear profiles is very limited.
For some exceptions, see \cite{jeong2007data}, \cite{paynabar2013monitoring},
\cite{grasso2014profile}, and \cite{paynabar2015change}.
There are two main challenges when monitoring multichannel profiles. The first one is that profiles are high-dimensional functions with intrinsic inner- and inter-channel correlations, and one needs to develop a dimension reduction method that can deal with such intricate correlations.
The second, probably more fundamental, challenge is that the functional structure of multi-channel profiles might change over time, and thus
the dimension reduction method should  be able to take into account the potential unknown change. 


The primary goal of this paper is to develop an effective statistical method for monitoring multichannel profiles.  Our methodology is inspired by the functional
Principal Component Analysis (PCA), which has been successfully applied by \cite{paynabar2013monitoring},
\cite{grasso2014profile}, and \cite{paynabar2015change}  to deal with intrinsic inner- and inter-channel correlations of profiles.
These existing methods follow the standard PCA approach to select a few principal components (projections or
eigenvectors) that contain a large amount of variation or information
in the profile data under the normal operational (in-control) state. This kind of dimension reduction
approach might be reasonable  from the estimation
or curve fitting/smoothing viewpoint under the normal operation state,
but unfortunately it is ineffective in the context of process monitoring,
especially for multivariate or multichannel profiles. This is because it does not reflect the possible change or fail to capture
the profile information under the out-of-control state. Here we propose to develop a PCA method that can automatically take into account the potential unknown change.


Note that there are two different phases of profile monitoring:
one is Phase I for offline analysis when a retrospective data set is used to estimate and refine the underlying model and its parameters, and the other is Phase II when the estimated model in Phase I is used for online process monitoring.
Here we focus on the Phase I analysis, and hopefully our results can shed new light
for Phase II monitoring of multichannel profiles as well.
In addition, we should acknowledge that the importance of dimension reduction and
feature selection for high-dimensional data via thresholding or shrinkage is well-known in
modern statistics, including the profile monitoring literature.
\cite{jeong2006wavelet} incorporated the hard thresholding
into the Hotelling $T^{2}$ statistics in the context of online monitoring
of single profiles, and \cite{jeong2007data} proposed
a hard thresholding method to obtain projection information by optimizing
``overall relative reconstruction error''.    
\cite{zou2012lasso}
applied LASSO shrinkage in linear model coefficients for online monitoring
linear profiles problem. However, these existing methods
use thresholding or shrinkage to conduct {\it one-shot} dimension reduction,
whereas our proposed methodology splits the dimension reduction process into two steps using two different methods:
PCA for the normal operation or in-control state, and soft-thresholding for the out-of-control state.

%

The remainder of this paper is organized as follows. In Section 2, we present
the mathematical formulation of multichannel profile monitoring. In Section 3, we propose our
thresholded PCA method, and provide a guideline on how to select the corresponding tuning
parameters. In Section 4, we use the real forging process data and simulations
to illustrate the usefulness of our proposed thresholded PCA  method.
Concluding remarks and future research directions are presented in Section 5.

\section{Problem Formulation and Background}

Suppose that a random sample of $m$ multichannel  profiles, each with $p$ channels, is collected from a production process. Mathematically, each of the $m$ multichannel  profile observations is a $p$-dimensional curve denoted by $\X_i(t) = (X_i^{(1)}(t),..., X_i^{(p)}(t))^T,$ where $t\in [0,1],$ for $i=1, \cdots, m.$ We assume that the process is initially in-control and at some unknown time $\tau,$ the process may become out-of-control in the sense of the mean shifts of the profiles $\X_{i}(t)$'s. Specifically, we assume that the data are from the change-point additive noise model
\begin{eqnarray} \label{eqn01}
\X_i(t)  =  \left\{
\begin{array}{ll}
  \bfmu_1 (t) + \Y_i(t),  & \hbox{when $i = 1,..., \tau,$} \\
  \bfmu_2 (t) + \Y_i(t), & \hbox{when $i = \tau+1, ..., m,$}
\end{array}
\right. \qquad \mbox{ for $0 \le t \le 1,$}
\end{eqnarray}
for some unknown $0 \le \tau < m,$ where  the $\Y_i(t)$'s are independent and identically distributed (i.i.d.) $p$-dimensional
``noise" curves with mean ${\bf 0,}$ i.e., $\Y_i(t)=(\Y_i^{(1)}(t), \cdots,  \Y_i^{(p)}(t))^T$ and $\E(\Y_i^{(j)}(t)) = 0$ for all dimension $j=1, \cdots, p$ and for all observations $i=1, \cdots, m.$

In Phase I profile monitoring,
$\bfmu_1 (t)$ and $\bfmu_2 (t)$ are two unknown $p$-dimensional mean functions, and we want to utilize the observed $\X_{i}(t)$'s to test the null hypothesis $H_0: \bfmu_1 (t) = \bfmu_2 (t)$ (i.e., no change)
against the alternative hypothesis $H_a: \bfmu_1 (t) \neq \bfmu_2 (t)$ (i.e., a change occurs at some unknown time $0 \le \tau  < m$). In addition, we also impose the classical  Type I probability error constraint
$P_{H_0}(\mbox{reject $H_0: \bfmu_1 (t) = \bfmu_2 (t)$})  \le \alpha,$
for some pre-specified constant $\alpha,$ e.g., $\alpha = 5\%.$

To test the hypothesis $H_0: \bfmu_1 (t) = \bfmu_2 (t)$  under  model (\ref{eqn01}) subject to the Type I error constraint,
it is important to make suitable assumptions of the correlation of both within and between profile channels. To characterize these correlations, as in \cite{paynabar2015change}, we apply Karhunen-Loeve expansion theorem to the $p$-dimensional noise curves $\Y_i(t)$:  there exists a set of orthonormal (orthogonal and unit norm) basis functions ${\cal V}=\{ v_k (t) \in L_2[0,1], k=1,2,...\},$ such that
\begin{eqnarray} \label{pca:eq1}
\Y_i(t) = \sum_{k \in {\cal V}}  \bfxi_{ik} v_k (t), \quad \mbox{ for } i=1,...,m,
\end{eqnarray}
where the number of elements of ${\cal V}$ could be either finite or infinite, and the coefficient $ \bfxi_{ik}=(\xi_{ik1}, \cdots, \xi_{ikp})$ is a $p$-dimensional vector. The key assumption we made is that the coefficients $\{\bfxi_{ik}\}$'s are i.i.d. $p$-dimensional random vectors with mean ${\bf 0}$ and covariance matrix $\Sigma_k$ over all $i=1, \cdots, m$ data points for each base $k \in {\cal V}$. Under this assumption, it is evident from (\ref{pca:eq1}) that the $p \times p$ covariance matrix $\Sigma_k$ satisfies
\begin{eqnarray} \label{eqn03}
\Sigma_k  = \E(\bfxi_{ik}\bfxi_{ik}^T) = \E \{ \int_{0}^{1} \Y_i(t)v_k(t) dt  \int_{0}^{1}  \Y_i(t)^T v_k(t) dt\},
\end{eqnarray}
since the basis functions $v_k(t)$'s are orthonormal for each $k \in {\cal V}$.

It is useful to briefly discuss the effect of (\ref{pca:eq1}) on the correlations of multichannel profiles. As in the standard functional data analysis, the real-valued basis  functions $v_k (t)$'s are closely related to the {\it inner-channel} correlation of the profiles. Meanwhile, since the $p$-dimensional curve is decomposed  into the same real-valued basis  functions $v_k (t)$'s in (\ref{pca:eq1}), the {\it inter-channel} correlations of the $p$-channel profiles are characterized by the correlation matrices
$\Sigma_k$'s in (\ref{eqn03}) of the coefficients $\{\bfxi_{ik}\}$'s. In practice, both the basis functions $v_k (t)$'s and the covariance matrices $\Sigma_k$'s are unknown and needed to be estimated, see the next section.


\section{Our Proposed Thresholded PCA  Methodology}

In this section, we develop a thresholded multivariate functional PCA methodology for Phase I monitoring of multichannel profiles.
For the purpose of easy understanding, this section is subdivided into three subsections. In Subsection 3.1, we review the multivariate functional PCA method that estimates the basis $v_k (t)$'s in (\ref{pca:eq1}) and the covariance matrices $\Sigma_k$'s in (\ref{eqn03}).
This allows us to reduce the data from the space of $p$-dimensional profiles $\X_i(t)$'s to the space of the coefficients  $\bfxi_{ik}$'s in (\ref{pca:eq1}) under the normal operational or in-control state.
In Subsection 3.2,   our proposed method is developed as a hypothesis test for the change-point model in (\ref{eqn01}) augmented by soft-thresholding technique that has a nature semi-Bayesian interpretation and is closely related to the generalized likelihood ratio test. Here the soft-thresholding selects significant coefficients $\bfxi_{ik}$'s in (\ref{pca:eq1}) that are likely affected by the change, and thus can be thought of as a further dimension reduction under the out-of-control state. In Subsection 3.3, based on asymptotic analysis, we provide a guidance on the choice of tuning parameters in our proposed thresholded PCA methodology.


\subsection{Basis and Covariance Estimation}

To have a better understanding of the basis and covariance matrix estimation under the change-point model in (\ref{eqn01}), we first consider the estimation under the unrealistic case when the noise functions $\Y_{i}(t)$'s in  (\ref{pca:eq1}) were observable.
Recall that the $p$-dimensional functions $\Y_{i}(t)$'s are decomposed into the same real-valued basis  functions $v_k (t)$'s in (\ref{pca:eq1}),
this motivates us to evaluate the {\it inner-channel} correlation of $\Y_{i}(t)$'s by the following covariance function:
\begin{eqnarray} \label{intro:eq1}
c(t,s)=\Cov \{\Y_i(t), \Y_i(s)\} 
= \sum_{j=1}^{p} \E(Y_i^{(j)}(t) \cdot Y_i^{(j)}(s)) \qquad \mbox{ for $0 \le t, s \le 1,$}
\end{eqnarray}
since $\Y_{i}(t)$ is a $p$-dimensional function with mean ${\bf 0}.$ When $p=1,$ the covariance function $c(t,s)$ in (\ref{intro:eq1}) is
well studied, and it is well-known that the bases $v_k(t)$'s are the eigenfunctions of  $c(t,s).$ Below we will show that similar conclusions also hold under our definition of the covariance function $c(t,s)$ in (\ref{intro:eq1}) for the general $p \ge 2$ case.

To see this, since the basis functions $v_{k}(t)$'s are  orthonormal, it follows from (\ref{pca:eq1}) that
$c(t,s) = \sum_{k=1}^{\infty} \sum_{j=1}^{p} \E[\xi_{ikj}^2] v_k(t) v_k(s),$
and
$\int_{0}^{1} c(t,s)v_k(s) ds = \lambda_k v_k(t),$
where $\lambda_{k}= \sum_{j=1}^{p} \E[\xi_{ikj}^2],$ and $\xi_{ikj}$ is the $j$-th component of the $p$-dimensional random vector $\bfxi_{ik}$ for $j=1, \cdots, p.$ Hence, the basis $v_k(t)$'s are the eigenfunctions of  $c(t,s)$ for any dimension $p \ge 2.$


It suffices to estimate the covariance function $c(t,s)$  in (\ref{intro:eq1})  from the observable profiles $\X_{i}(t).$ While the noise terms $\Y_i(t)$'s are unobservable, a good news of the  change-point additive noise model in (\ref{eqn01}) is that the differences $\Y_{i+1}(t) - \Y_i(t) = \X_{i+1}(t) - \X_i(t)$ are observable for   all $1 \le i \le m-1$ except $i = \tau$ (the change-point). Thus the covariance function $c(t,s)$ in (\ref{intro:eq1}) can be estimated by $\Y_{i+1}(t) - \Y_i(t),$ which yields the approximation:
\begin{eqnarray} \label{pca:c}
\hat{c}(t,s) = \frac{1}{2(m-1)} \sum_{i= 1}^{m-1} (\X_{i+1}(t) - \X_i(t))^T (\X_{i+1}(s) - \X_i(s)).
\end{eqnarray}
Note that the denominator is $2(m-1),$ and since the $\Y_{i}(t)$'s are i.i.d. over $i=1, \cdots,m,$ the estimated function $\hat{c}(t,s)$ in (\ref{pca:c}) is consistent under the reasonable regularity assumption of the alternative hypothesis, see Remark \#2 in \cite{paynabar2015change}.

Next, the estimates of basis functions $\hat{v}_k (t)$'s can be found as the eigenfunctions of  $\hat{c}(t,s)$ in (\ref{pca:c}). As for the estimation of  the covariance matrix $\Sigma_k$ in  (\ref{eqn03}) of coefficients $\bfxi_{ik},$  we again take advantage of the differences $\Y_{i+1}(t) - \Y_i(t)$ under the change-point additive noise model in (\ref{eqn01}), and approximate it  by
\begin{eqnarray} \label{pca:sigma}
\hat{\Sigma}_k  = \frac{1}{2(m-1)} \sum_{i=1}^{m-1} \int_{0}^{1} \{\X_{i+1} (t) - \X_{i} (t) \}\hat{v}_k(t) dt \int_{0}^{1} \{\X_{i+1} (t) - \X_{i} (t) \}^T \hat{v}_k(t) dt.
\end{eqnarray}

We follow the standard PCA literature to focus on the first $d$ largest eigenvalues  of the function $\hat{c}(t,s)$ in (\ref{pca:c}), and consider the corresponding $d$ eigenfunctions $\hat{v}_k (t)$'s.  However, our choice of  the actual value of $d$ will be different here. From the dimension reduction viewpoint,  the standard PCA methods often reduce the data directly to a low-dimensional space, and thus the value of $d$ is often chosen to  be relatively small.
Meanwhile, for our proposed method,  the dimension reduction process is split into two steps that correspond to the normal operation state and the out-of-control state, respectively. The PCA is used only in the first step to reduce the data from the infinitely functional (or super-high-dimensional) space to an intermediate space of $R^{d},$ which will be further reduced to a lower-dimensional space in the second step. As a result, the number  $d$  of the chosen principal components of the PCA  can be moderately large for our proposed method, e.g., fifties or hundreds.

\subsection{Thresholded PCA for Monitoring}

We are ready to present our proposed method that utilizes the observed profiles $\X_i(t)$'s to test $H_0: \bfmu_1 (t) = \bfmu_2 (t)$  under the change-point additive noise model (\ref{eqn01}).
Intuitively, it is natural to construct a test statistic based on the estimation of $\bfmu_1 (t) - \bfmu_2 (t).$ This suggests us to compare the difference of profile sample means before and after a potential change-point $\ell = 1, 2,...,m-1,$
\begin{eqnarray} \label{know:delta}
{\bf \Delta}_{\ell}(t)  = \sqrt{\frac{\ell(m-\ell)}{m}} \left\{ \frac{1}{\ell}\sum_{i = 1}^{\ell} \X_i(t) -\frac{1}{m-\ell} \sum_{i=\ell+1}^{m}  \X_i(t)\right\}.
\end{eqnarray}
Here the term $\sqrt{\ell(m-\ell)/m}$  scales the difference and standardizes the variance of profile difference. Note that the function ${\bf \Delta}_{\ell}(t)$ in (\ref{know:delta}) would have mean ${\bf 0}$ when $H_0: \bfmu_1 (t) = \bfmu_2 (t)$ is true, but have non-zero mean under $H_a: \bfmu_1 (t) \ne \bfmu_2 (t)$ when $\ell = \tau$ (the change-point). 

Next, with the estimated orthonormal basis $\hat v_k(t)$'s and estimated covariance matrix $\hat{\Sigma}_k$ in (\ref{pca:sigma}),
we apply the PCA decomposition in (\ref{pca:eq1}) to the function ${\bf \Delta}_{\ell}(t)$ in (\ref{know:delta}). This essentially projects the test statistics from the functional space to a $d$-dimensional space under the normal operational or in-control state.
Specifically,
for each candidate change-point $\ell = 1, 2,...,m-1,$ define the projection to  each of the first $d$ principal components,
$\bfeta_{\ell k} = \int_{0}^{1} {\bf \Delta}_{\ell}(t) \hat{v}_k(t)dt,$
and then compute the corresponding real-valued statistic
\begin{eqnarray}  \label{know:eta2}
U_{\ell, k} = \bfeta_{\ell k}^T \hat{\Sigma}_k^{-1}\bfeta_{\ell k}
\end{eqnarray}
for $k=1, 2, \cdots, d,$ where $\hat{\Sigma}_k$ is defined in (\ref{pca:sigma}).

Note that the statistics $U_{\ell, k}$'s in (\ref{know:eta2}) are motivated from the scenario when the basis $\nu_k(t)$ and $\Sigma_k$ are known:
if the estimates $\hat{v}_k(t)$ and $\hat{\Sigma}_k$ are replaced by their true values, it is straightforward from (\ref{eqn01}) to show that $\bfeta_{\ell k} \sim N({\bf 0}, \Sigma_k)$ under the null hypothesis $H_0: \bfmu_1 (t) = \bfmu_2 (t)$  but $\bfeta_{\ell k} \sim N(\int_{0}^{1} {\bf \Delta}_{\ell}(t) v_k(t)dt, \Sigma_k)$ under the alternative hypothesis $H_a: \bfmu_1 (t) \ne \bfmu_2 (t).$ Hence, when the basis $\nu_k(t)$ and $\Sigma_k$ are known, the $U_{\ell, k}$'s in (\ref{know:eta2}) are $\chi_{p}^2$-distributed under $H_0$ but should be stochastically larger than $\chi_{p}^2$ under $H_a.$  When the estimates $\hat{v}_k(t)$ and $\hat{\Sigma}_k$ are used,  we expect that similar conclusions also hold approximately, e.g., whether the value of $U_{\ell, k}$ in (\ref{know:eta2})  is large or small indicates whether there is a change along the principal component $\hat{v}_k(t)$ or not.

Finally, our proposed thresholded PCA methodology  considers the soft-thresholding  transformation of the $U_{\ell, k}$'s in (\ref{know:eta2}), so as to smooth out those noisy principal component $\hat{v}_k(t)$'s that do not provide information about the change under the out-of-control state. To be more rigorous, we propose a test statistic defined by
\begin{eqnarray} \label{pca:Q}
Q_m =  \max_{1 \le \ell < m}\sum_{k=1}^{d}(U_{\ell, k}  - c)^+,
\end{eqnarray}
for some pre-specified ``soft-thresholding" parameter $c \ge 0.$ Here $(u-c)^{+} = \max(u -c, 0).$ Then we reject the null hypothesis $H_0: \bfmu_1 (t) = \bfmu_2 (t)$ if and only if
\begin{eqnarray} \label{know:Q}
Q_{m} > L
\end{eqnarray}
for some pre-determined threshold $L.$ The choices of the constants $c$ and $L$ will be discussed in more detail in the next section. When $Q_m > L,$ we not only claim that there exists a change point, but also can estimate the change point by
\begin{eqnarray} \label{pca:tau}
\hat{\tau} = \arg \max_{1 \le \ell < m}\sum_{k=1}^{d} (U_{\ell, k}  - c)^+.
\end{eqnarray}

It is informative to provide some high-level insights of the test statistic  $Q_m$ in (\ref{pca:Q}). Since we do not know the true change-point $\tau,$ it is natural to maximize  (\ref{pca:Q})  over all candidate change-points $\tau = \ell$ for $1 \le \ell < m$  from the maximum likelihood estimation or generalized likelihood ratio test viewpoints. The summation of the soft-thresholding transformation $(U_{\ell, k}  - c)^+$  in  (\ref{pca:Q}) is more fundamental and  can be interpreted from the following semi-Bayesian viewpoint. For a given candidate change-point $\ell,$ let $Z_{k}$ be the indicator whether the $k$-th principal component is affected by the change in the out-of-control state or not, for $k=1, \ldots, d.$ Assume that  all principal components are independent, and each has a prior probability $\pi$ getting affected by the changing event. That is, assume that the changing indicators $Z_1, \ldots, Z_{d}$ are iid with probability mass function $\Prob(Z_{k} = 1) = \pi = 1- \Prob(Z_{k} = 0).$  When $Z_{k} = 1,$ the $k$-th principal component is affected, and $U_{\ell, k}$ in (\ref{know:eta2}) represents the evidence of possible change in the log-likelihood-ratio scale. Treating $Z_{k}$'s as the hidden states, and then the joint log-likelihood ratio statistic of $Z_{k}$'s and $X_{k,n}$ when testing $H_0: Z_1 = \ldots = Z_{d} = 0$ (no change) is
\begin{eqnarray*}
LLR(n) = \sum_{k=1}^{d} \{ Z_{k} (\log \pi + U_{\ell, k}) + (1-Z_{k}) \log (1-\pi) \}  - \sum_{k=1}^{d} \log (1-\pi),
\end{eqnarray*}
which becomes $\sum_{k=1}^{d} Z_{k} \{U_{\ell, k}  - \log((1-\pi) /\pi) \}.$
Since the $Z_{k}$'s are unobservable, it is natural to maximize $LLR(n)$ over $Z_{1}, \ldots, Z_{d} \in \{0,1\}.$
Hence, the generalized log-likelihood ratio becomes
$\sum_{k=1}^{d} \max\{U_{\ell, k}  - \log((1-\pi)/\pi), 0 \},$
which is exactly  our test statistic $Q_{m}$ in (\ref{pca:Q}). 

We should acknowledge that from the mathematical viewpoint, the multivariate functional PCA-based monitoring method in  \cite{paynabar2015change} is the special case of $Q_{m}$ in (\ref{pca:Q}) when the soft-thresholding parameter $c = 0,$ which is reasonable in that context because the number $d$ of principal components is small (e.g., $d=15$).  However, our proposed method is a non-trivial extension of
\cite{paynabar2015change} from the statistical or dimension reduction  viewpoint: we consider a moderately large value $d$ of principal components (e.g., $d= 45$), and  a suitable choice of the soft-thresholding parameter $c > 0$  in (\ref{pca:Q}) is essential to conduct another level of dimension reduction to smooth out those  principal components that do not provide information of the change under the out-of-control state.

\subsection{The Choices of Tuning Parameters}\label{sec: tuning_par}

There are two tuning parameters in our proposed  thresholded PCA methodology based on the test statistic  $Q_m$ in (\ref{pca:Q}): one is the soft-thresholding parameter $c$ in (\ref{pca:Q}), and the other is the threshold $L$ in (\ref{know:Q}). Practically, one needs to determine $c$ first before selecting $L,$ but below we will present the choice of $L$ first for a given $c$ since it is easier to understand from the statistical viewpoint.

In order to find the threshold $L$ for our proposed methodology to satisfy the Type I error probability constraint,
assume, for now, that the constant $c$ in (\ref{pca:Q}) is given. Then the constraint
becomes $\Prob_{H_0}(Q_m >L) \le \alpha.$ Hence, the threshold $L$ should be   the upper $\alpha$ quantile of the distribution of $Q_{m}$  in (\ref{pca:Q}) for a given $c$  under $H_0,$ which can be simulated by Monte Carlo method based on normal profiles and models, see \cite{paynabar2015change}.



Let us now discuss the choice of soft-thresholding parameter $c$ in (\ref{pca:Q}). The baseline choice of $c$ is $c_0 = 0,$ which yields the approach of  \cite{paynabar2015change}  for the scenario when the number $d$ of selected principal components is small. Intuitively, when the number $d$ of principal components are large, the soft-thresholding parameter $c > 0$ in (\ref{pca:Q}) should be large enough to filter out those non-changing bases $\hat{v}_{k}(t)$'s, but cannot be too large to remove some changing principal components and lower the signal-to-noise ratios.  Hence, a suitable choice of $c$ will depend on the specific   $H_{a}$ and its effects on the basis projections.

Below we will discuss two different heuristic choices of the soft-thresholding parameter $c > 0.$ For that purpose, by (\ref{pca:Q}), we have
\begin{eqnarray} \label{eq:Lc0}
\Prob(\sum_{k=1}^{d} (U_{\ell, k}-c)^+>L) \le \Prob(Q_{m} > L)  &\le&
\sum_{\ell=1}^{m-1} \Prob(\sum_{k=1}^{d} (U_{\ell, k}-c)^+>L),
\end{eqnarray}
which becomes $(m-1) \Prob(\sum_{k=1}^{d} (U_{\ell, k}-c)^+>L),$ as the data are iid over $\ell=1,\cdots,m-1.$ Hence, from the asymptotic viewpoint, $\Prob(Q_{m} > L)$ and $\Prob(\sum_{k=1}^{d} (U_{\ell, k}-c)^+>L)$ go to $0$ at the same rate when $m$ is fixed. In particular, when the Type I error constraint $\alpha$
goes to $0,$ the main probability of interest is to estimate
 $\Prob_{H_0}(\sum_{k=1}^{d} (U_{\ell, k}-c)^+>L_c),$
where $L_{c}$ is chosen so that this probably $\le \alpha.$ Our proposed choices of $c$ correspond to two different methods to approximate the distribution of $\sum_{k=1}^{d} (U_{\ell, k}-c)^+$ under $H_0:$ one is the central limit theorem (CLT) when $c$ is small, and the other is the extreme theorem when $c$ is large. Since these two methods yield different results on $c$, we present them separately in Proposition 1, which assumes that $\chi_p^2$ approximation applies to $U_{\ell, k}$'s.


\begin{proposition} Assume that $U_{\ell, k} \sim \chi_p^2$ under $H_0,$ 
for all $k = 1,...,d;$
\begin{description}
  \item [(a)] (The CLT approximation when $c$ is small). Assume further that under $H_{a}$, exactly $d_0$ out of $d$ principal components are affected in the sense that
$U_{\ell, k} \sim \chi_p^2(\delta^2 p) =  \bfepsilon_{\ell k}^T \bfepsilon_{\ell k}$ with $\bfepsilon_{\ell k} \sim N(\delta, I_p)$ for $k = 1,...,d_0,$ and $U_{\ell, k} \sim \chi_p^2$  for $k=d_0+1,...,d.$  Then when both $d_0$ and $d -d_0$ are large, an appropriate choice of $c$ is
 \begin{eqnarray} \label{eq:c1}
c_1 = \mbox{arg min}_{c \ge 0}  \left\{-\frac{(\mu_c^{(1)} - \mu_c)d_0}{\sqrt{d_0 (\sigma_c^{(1)})^2 + (d-d_0) (\sigma_c)^2}} + \frac{\sqrt{d}\sigma_c}{\sqrt{d_0 (\sigma_c^{(1)})^2 +
(d-d_0) (\sigma_c)^2}} z_{\alpha}\right\}, \quad
\end{eqnarray}
where $\mu_c = \E_0(U_{\ell, k}-c)^+$ and $\sigma_c   = \text{Var}_0(U_{\ell, k}-c)^+$ when $U_{\ell, k} \sim \chi_p^2;$
$\mu^{(1)}_c = \E_1(U_{\ell, k}-c)^+$ and $\sigma^{(1)}_c   = \text{Var}_1(U_{\ell, k}-c)^+$ when $U_{\ell, k} \sim \chi_p^2(\delta^2 p).$

  \item [(b)] (The extreme theory approximation when $c$ is large). For fixed $p$ channels, as $d \rightarrow \infty,$ the soft-thresholding parameter $c$ can be chosen as 
\begin{eqnarray} \label{eq:c2}
c_2 \approx p + 2 \log(d).
\end{eqnarray}
\end{description}
\end{proposition}

{\bf Proof:} Due to page limit, let us only provide a sketch of the proof. In part (a), the $c_1$ value  maximizes the power of the test under $H_a$ subject to the Type I error constraint $\alpha,$ and the CLT is used to approximate error probabilities. That is the reason why we need some prior information  on $d_0$ and $\delta$ under $H_a.$ In our numerical studies, when such prior information of $H_a$ is not available, our  experiences suggest that $\delta = 1$ and $d_0 = d/3$ yield a good robust result under our simulation numerical setting.

The rationale of part (b) is completely different, and is similar to use the following well-known fact to choose the soft-thresholding parameter of $\sqrt{2 \log(d)}$ for $d$ iid $N(0,1)$ random variables, see \cite{fan1996test},
\[
\lim_{d \rightarrow \infty} \frac{\max_{1 \le k \le d} |Z_{k}|}{\sqrt{2 \log(d)}} = 1 \quad \mbox{almost surely}
\]
when the $Z_{k}$'s are iid $N(0,1).$  Here  we extend the critical value  from $\sqrt{2 \log(d)}$  for the $N(0,1)$-distributed $Z_{k}$'s to $c_2$ for the $\chi_{p}^2$-distributed $U_{\ell, k}$'s for fixed $p$ as $d \rightarrow \infty.$ These two critical values are asymptotically equivalent when $p=1,$ as $N(0,1)^2$ is $\chi_{p}^2$-distributed with $p=1.$
To prove part (b) rigorously, we first use the fact
\begin{eqnarray*}
\Prob_{H_0}(\sum_{k=1}^{d} (U_{\ell, k}-c)^+>L_c)
 < \Prob_{H_0}\Big(\max_{1 \le k \le d} U_{\ell, k} > c \Big) <  \sum_{k=1}^{d} \Prob_{H_0}\Big( U_{\ell, k} > c \Big) = d \Prob\big( \chi_{p}^{2} > c\big),
\end{eqnarray*}
since we assume $U_{\ell, k} \sim \chi_{p}^{2}$ under $H_0.$  Next, we need use the asymptotic expression of $\Prob ( \chi_{p}^{2} > c)$ in \cite{inglot2006asymptotic} to establish a useful lemma that $\log \Prob( \chi_{p}^{2} > c) = -(c-p)/2 + O(\log c)$ for fixed $p$ as $c \rightarrow \infty.$ Then it is straightforward to prove part (b).


\section{Case Study}

In this section, we apply our proposed thresholded PCA method to the real forging manufacturing process dataset in Figures \ref{fig:introa} and \ref{fig:introb} in the Introduction. This dataset includes 207 normal profiles under the in-control state  and 69 different fault profiles under the out-of-control state.  It was analyzed in \cite{paynabar2015change} whose method can be thought of as the special case of our proposed method with the specific soft-thresholding parameter $c_0=0.$ Below the choice of $c_0=0$ will be regarded as the baseline method, and we will focus on whether the values of $c_1$ and $c_2$ in Proposition 1 for the soft-thresholding parameter $c$ in (\ref{pca:Q}) can improve the performance or not as compared to the baseline value $c_0=0.$


First, we consider a specific case study  setting in \cite{paynabar2015change} where 207 normal profiles are followed by the $69$ fault profiles, i.e., the change-point $\tau = 207$ for the change-point   model in (\ref{eqn01}), and  the baseline method $c_0=0$ can successfully detect the true change-point. Our experiences show that our proposed  method with either $c_1$ or $c_2$   can also correctly detect the change-point.
In other words, if the change is significantly large, then all reasonable profile monitoring algorithms, including our proposed methods with any of the three $c$ values in (\ref{pca:Q}), will be able to detect the change correctly.

Below we will conduct extensive simulation studies that focus on detecting smaller changes. 
For better presentation, the remainder of this section is divided into two subsections. In subsection 4.1, we use the real profiles  and B-splines to present the generative models of profiles under the in-control state and $2\times3\times 7 =42$ different out-of-control states. This allows us to generate observed profiles $\X_{i}(t)$'s from the change-point additive noise model in (\ref{eqn01}).
In subsection 4.2, our proposed thresholded PCA methods are applied to the generated profiles $\X_{i}(t)$'s, and the performance of the values of $c_1$ and $c_2$ in Proposition 1 is then compared with that of the baseline value $c_0 = 0.$

\subsection{Profile Generative  Models}
\label{sec:fit_data}

Let us provide a high-level description of our simulation setting. In each run of our simulation studies below, we generate $m=200$ profiles from the change-point  model in (\ref{eqn01}) with change-point $\tau=100,$ i.e., the first 100 profiles, $\X_{1}(t),...,\X_{100}(t)$, are generated from the in-control state,
and the last $100$ profiles, $\X_{101}(t),...,\X_{200}(t)$ are generated from one of the $42$ out-of-control states.
For each  set of $m=200$ simulated profiles, our proposed thresholded PCA method with  three different soft-thresholding parameters
$c_{0},c_{1},c_{2},$ are applied to see whether they are able to correctly detect the change $\tau = 100$ or not.
This process is repeated for $200$ times, and the average performances  are reported and compared for three different parameters
$c_{0},c_{1},c_{2}.$  It is important to emphasize that the  generative models below are only used to generate the $m=200$ observable profiles  $\X_{i}(t)$'s. Our proposed thresholded PCA methods are applied to those $m=200$ profiles, and do not use any information or knowledge of these profile  generative models.

\begin{figure}
\centering 
\includegraphics[width=3in]{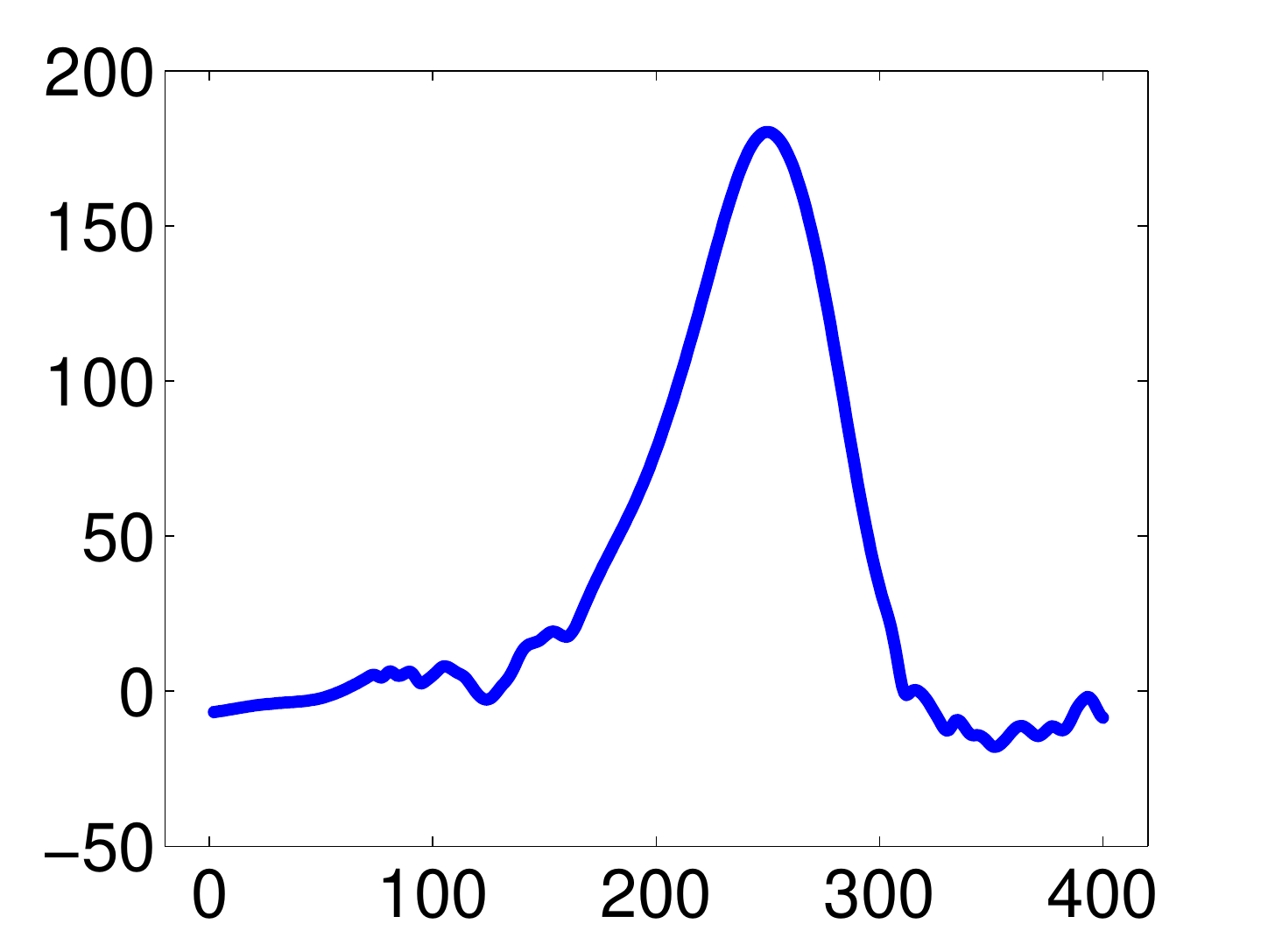}\\
 \protect\caption{This figure plots the simulated in-control single profile $\X^{(1)}_m(t)$ based on an average of 200 replications. Interval {[}0, 400{]} in the x-axis corresponds to $t \in  {[}0/400, 400/400{]}$. This plot shows that the generative model in (\ref{eqn019}) under the in-control state indeed produced profiles that mimic the profiles from real forging process dataset in Figures \ref{fig:introa} and \ref{fig:introb}.}
 \label{fig:simu_normal}
\end{figure}

For the generative models for profiles under  the in-control state,  we propose to build such a model
by applying B-splines to the  $207$ normal profiles, $\X_{1}(t),...,\X_{207}(t),$ in the real forging dataset. To be more specific, we generate an unevenly
spaced set of 66 B-spline basis in $[0,1],$ and after orthogonalization and normalization we obtain basis ${B_{1}(t),...,B_{66}(t)}$ using
the ``orthogonalsplinebasis'' Package in  the free statistical software R 3.1.2. Based on our experiences, the choice of $66$ basis yields the best tradeoff to balance the fitting of normal profiles and the computational simplicity, but it can easily be changed to another number.  Then our proposed generative model for normal profiles is of the form
\begin{eqnarray} \label{eqn019}
\X(t)=\sum_{i=1}^{66}\widetilde{\bftheta}_{i}B_{i}(t),
\end{eqnarray}
where the $4$-dimensional vectors  $\widetilde{\bftheta}_{i}$'s are assumed to be multivariate normally distributed with parameters estimated from the observed $207$ normal profiles,  see Figure \ref{fig:simu_normal}.

For profiles under the out-of-control (OC) state, we assume that the generative OC model is the same as (\ref{eqn019}) but the means of $\widetilde{\bftheta}_{i}$'s might change. We will consider a total of $2 \times 3 \times 7 = 42$ different  OC cases, depending on three different factors. First, we consider two different scenarios, depending on how many components/channels of the   $4$-dimensional random vector $(\widetilde{\theta}_{i}^{(1)}, \widetilde{\theta}_{i}^{(2)}, \widetilde{\theta}_{i}^{(3)}, \widetilde{\theta}_{i}^{(4)})$ are involved
with the change: (A) All $4$ components/channels have new OC mean; and (B) Only the first $2$ components/channels, $\widetilde{\theta}_{i}^{(1)}$ and $\widetilde{\theta}_{i}^{(2)}$ have OC mean (our proposed methods are not designed for Scenario B, and we run simulation to see their performance).
Second, we consider three cases, depending on which subset of the $66$ different $\widetilde{\bftheta}_{i}$ in the model (\ref{eqn019}) changes their means, or equivalently, which location or interval of $[0, 1]$ changes at the original profile scale:
(I) a local change for $30 \le i \le 37;$ (II) a local change for $16 \le i \le 29$ and (III) a global change for all $1 \le i \le 66.$ In the context of the original profiles, the locations of these three changes occur over intervals $\frac{200}{400} \le t \le \frac{300}{400}, \frac{99}{400} \le t \le \frac{149}{400}$ and $0 \le t \le 1,$ respectively. Finally,  we consider seven different magnitude values, so as to have reasonable detection powers regardless of the locations of the change. In particular, when the real-valued mean  of $\widetilde{\theta}_{i}^{(j)}$'s changes from $\theta_i$ to $\theta_i + 0.005 + 0.005*\Delta,$
we set $\Delta= h+1$ for local change in Case (I), $\Delta = h$ for local change in Case (II), and $\Delta = 0.1*h$ for global change in case (III). Here there are seven values of $h:$  $h=1,2, \cdots, 7.$ Note that given the same magnitude of the change, it is the most difficult to detect the local change of Case (I) (where the peak of the profile occurs), and it is the easiest to detect the global change of Case (III). Here we assign different magnitudes so that the detection powers of these cases are comparable.    In summary, there are  $2 \times 3 \times 7=42$ OC cases depending on the channel, location, and magnitude of the changes, and  all numerical values are inspired from the real forging dataset.

\subsection{Performance Comparison}

In this subsection, we report the performance of our proposed thresholded PCA method with three different choices of the soft-thresholding parameter $c$, and our objective is to see whether the $c_1$ and $c_2$ in Proposition 1 will yield a better performance as compared to the baseline $c_0 = 0$ in the sense of detecting those $2\times 3 \times 7 = 42$ OC cases.

In order to have a fair comparison, we fix the number of principal components as $d=45$ for all three choices of soft-thresholding $c$ values, since on average that will explain more than $90\%$ of the profiles variance. In addition, for each method, we choose the threshold $L$ in (\ref{know:Q}) to satisfy  Type I error constraint $\alpha = 0.05.$ Also our proposed methods were developed under the assumption that all $4$ components/channels are affected, and the magnitudes of the changes are unknown.  Table \ref{tab:val_c_unknown} lists the specific  values of $c_{0},c_{1},c_{2}$ used in our study. Note that the value of $c_0=0$ and $c_2$ do not depend on the location of the change, but the value of $c_1$ depends on the location of the change.

\begin{figure}
 \centering

     \includegraphics[width=2.1in]{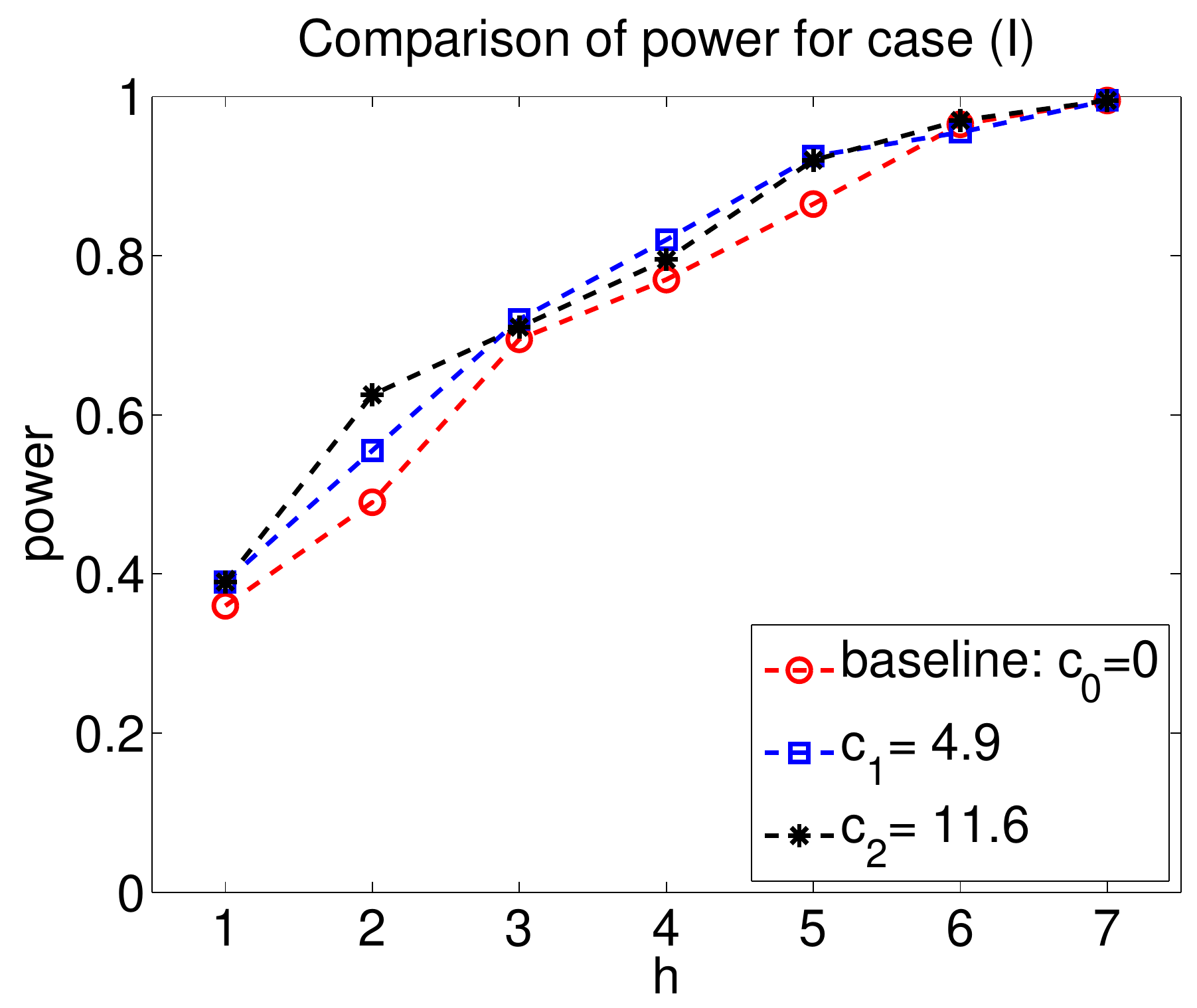}\\
    \includegraphics[width=2.1in]{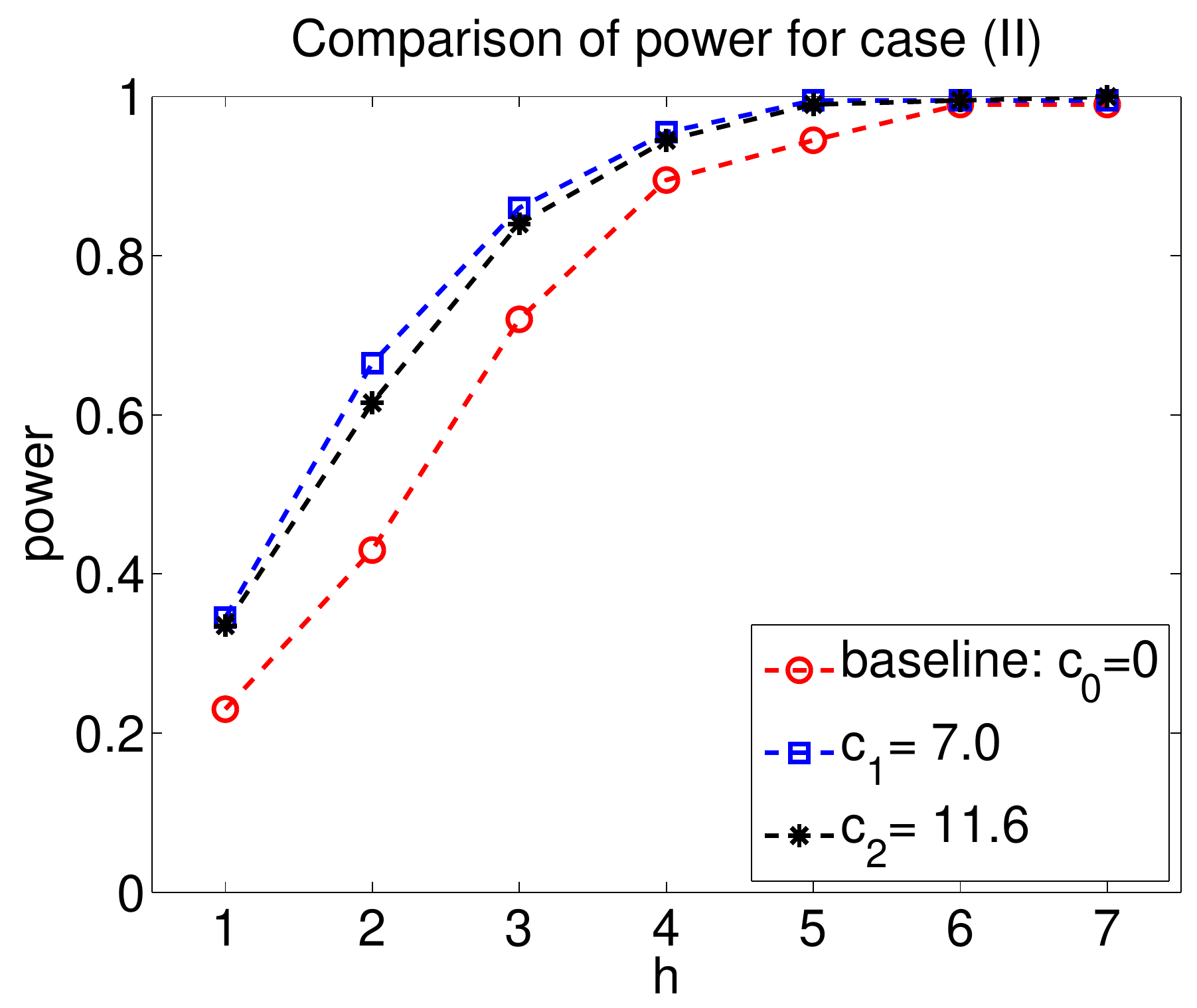}\\
   \includegraphics[width=2.1in]{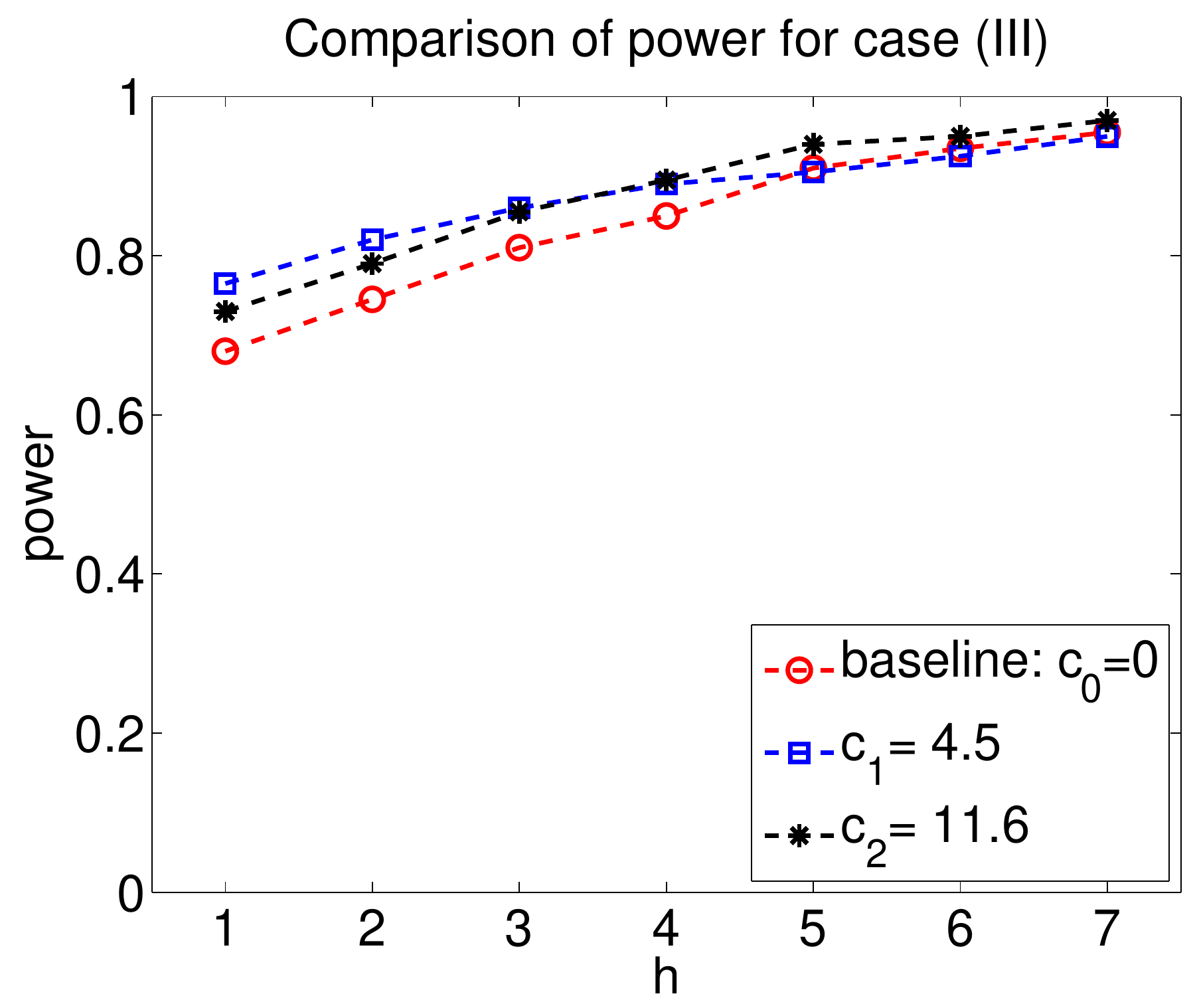}

 \caption{When all 4 channels/components are affected. The three plots correspond to three OC cases,  depending on which subset of the $66$ different $\widetilde{\bftheta}_{i}$ in the model (\ref{eqn019}) changes their means.
  {\it Upper}: case (I) with a local change for $30 \le i \le 37;$ {\it Medium}: case (II) with a local change for $16 \le i \le 29$ and  {\it Bottom}: case  (III) with a global change for all $1 \le i \le 66.$
  In each figure, each curve represents our proposed method with a specific soft-thresholding $c$ values: Red line with circle ($c_0$); blue line with square ($c_1$); and black line with star ($c_2$).  The detection  power of each method is plotted as the function of the $7$ different change magnitudes.
}
\label{fig:unknown_result}
\end{figure}

Figure \ref{fig:unknown_result} plots the detection power of our proposed methods
with three different choices of soft-thresholding $c$ values as functions of change magnitudes
when all $4$ components/channels of $\bftheta_{i}$ are actually changed simultaneously.
The top panel deals with the OC-case (I) where a local
change affects the rise, peak, and fall segments of the profiles, and all
three methods seem to have comparable detection powers, although $c_0=0$ is slightly worse.
The middle panel shows that under the OC case (II), both $c_{1}$
and $c_{2}$ can greatly improve the detection power as compared
with the baseline  $c_{0} = 0,$ especially when the change magnitude is small
(e.g., $h\leq5$).  For large change magnitudes, all three methods have detection power close to $1,$ implying that all reasonable methods
should be able to detect large changes.

A surprising observation of Figure \ref{fig:unknown_result} is the  bottom panel that considers the OC case (III) when a global
change occurs over  $[0,1].$  Intuitively, for a global change, one would expect that the change affects
all principal components and hence thresholding might not help. However, the bottom panel of Figure \ref{fig:unknown_result} is counter-intuitive, as both $c_{1}$ and $c_{2}$ seem to yield a  larger detection power than $c_{0} = 0$, especially for small   magnitude $h$.
To gain a deep understanding,
Figure \ref{fig:U_{lk}}  plots
the box plot of $U_{\ell=100,k}$ under the both IC and OC-case(III) states for all  $d=45$ principal components.
From the box plots, for the global change, it is surprising that almost half of $U_{\ell,k}$'s have a similar or smaller median value under OC than IC. We feel that this is the reason why soft-thresholding  help improve the detection power in the global change case, as it can filter out those $U_{\ell,k}$'s that have smaller OC values.


We also evaluate the performance of our proposed method in terms of estimating the change-point
$\tau$. When the true $\tau = 100$ is estimated as $\hat{\tau},$ we consider three different measures: $\E(|\hat{\tau} - \tau |),$ $P(|\tau-\hat{\tau}|\le1)$ (denoted
by P1) and $P(|\tau-\hat{\tau}|\le3)$ (denoted by P3). Table \ref{tab:result} reports the Monte Carlo simulation results under these three  criteria based on $200$ runs.
In general all three values $c_0, c_1$ and $c_2$ yield comparable results in terms of  estimating $\tau,$ and it is interesting to note that the thresholding values $c_1$ and $c_2$ often have larger P1 and P3 than the baseline $c_0 = 0$ for the OC case (II) with the local-mean shift cases. This suggests that thresholding might be able to locate the small, local change more precisely.
One ``strange" observation in Table \ref{tab:result} is that $\E(|\hat{\tau} - \tau |)$ is not necessarily monotone as a function of the change magnitude $h.$ We do not have a deep insight, and one possible explanation  is because $|\hat{\tau} - \tau|$ takes on the integer values, $0,1,2, \cdots,100,$ since both are integers.



\begin{figure}
 \centering

    \includegraphics[width=2.1in]{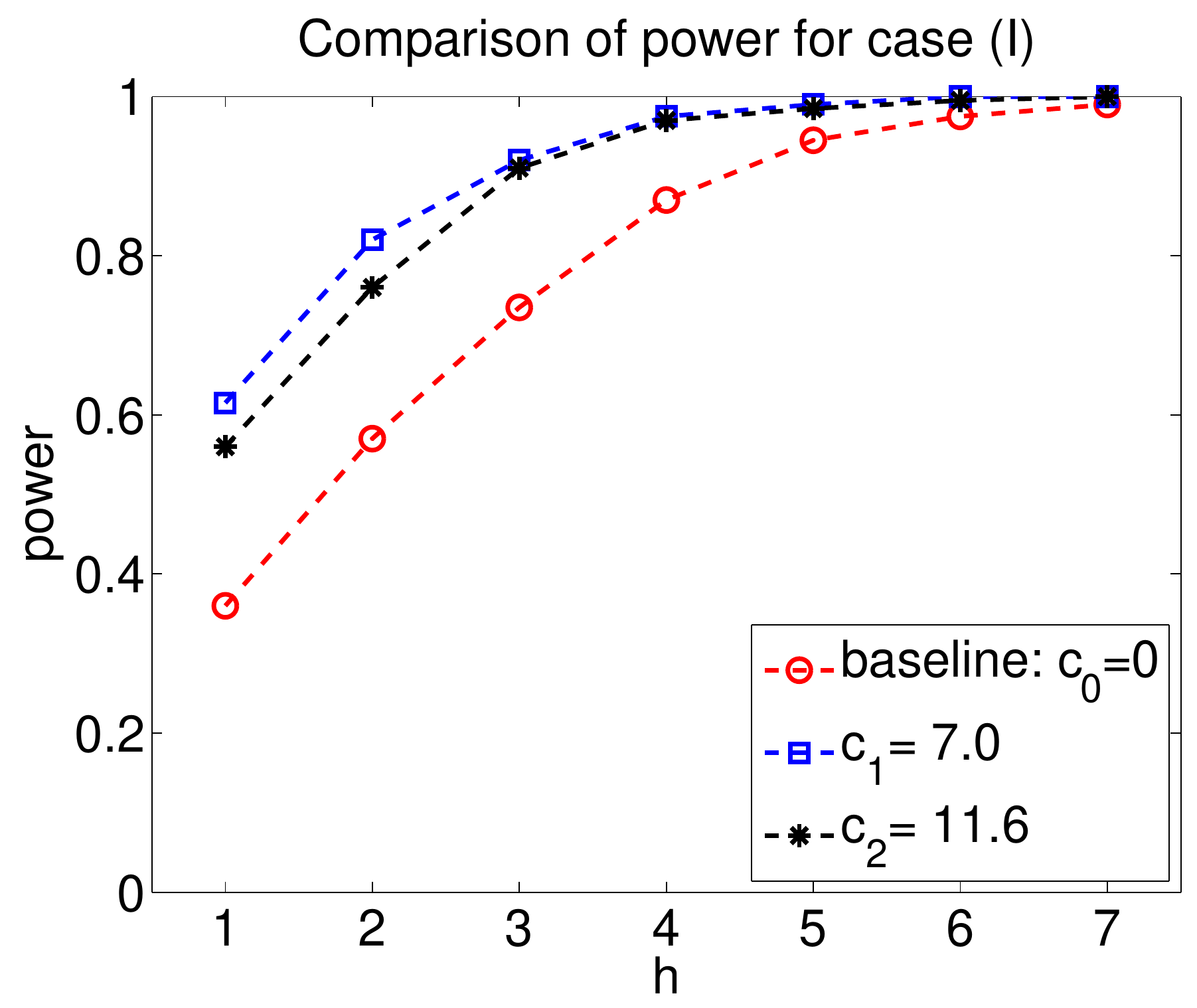} \\
    \includegraphics[width=2.1in]{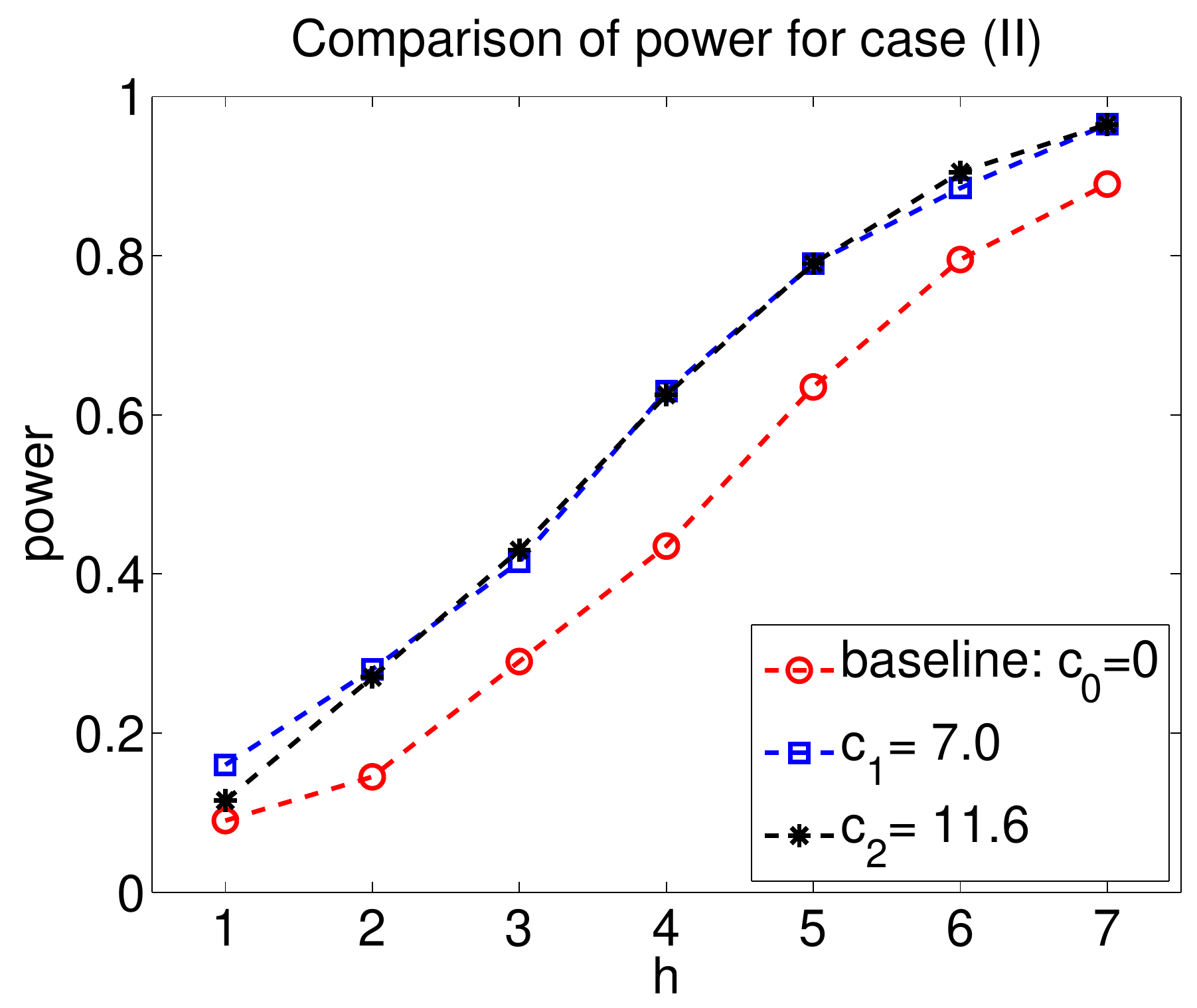}\\
  \includegraphics[width=2.1in]{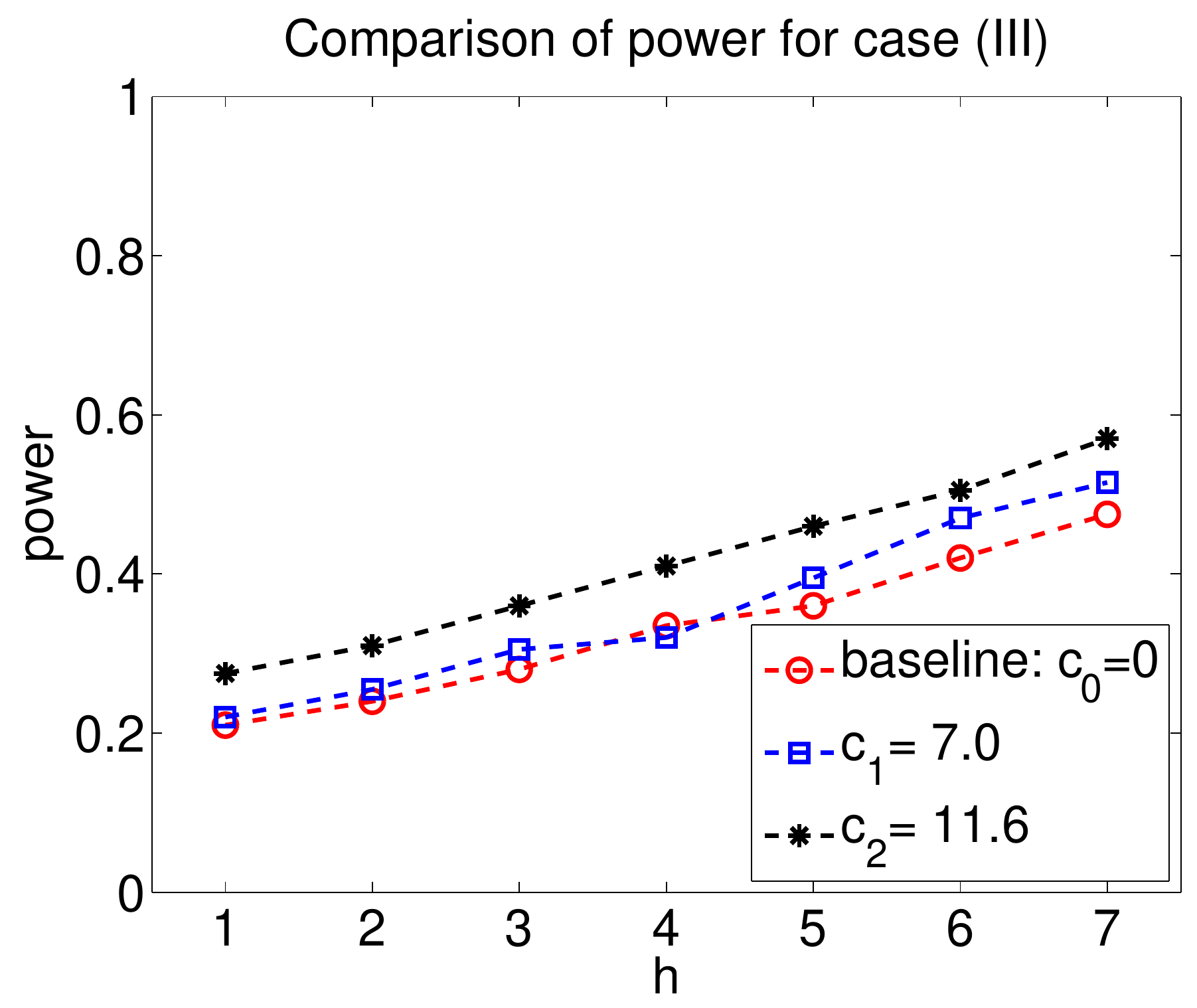}

 \caption{When only $2$ out of 4 channels/components are affected. The three plots correspond to three OC cases,  depending on which subset of the $66$ different $\widetilde{\bftheta}_{i}$ in the model (\ref{eqn019}) changes their means.
  {\it Upper}: case (I) with a local change for $30 \le i \le 37;$ {\it Medium}: case (II) with a local change for $16 \le i \le 29$ and  {\it Bottom}: case  (III) with a global change for all $1 \le i \le 66.$
  In each figure, each curve represents our proposed method with a specific soft-thresholding $c$ values: Red line with circle ($c_0$); blue line with square ($c_1$); and black line with star ($c_2$).  The detection  power of each method is plotted as the function of the $7$ different change magnitudes.
 }
\label{fig:unknown_result_2ch}
\end{figure}

Figure \ref{fig:unknown_result_2ch} plots the detection power of our proposed methods when only 2 out of 4 channels/components are affected.
It is clear from the top and middle panels of Figure \ref{fig:unknown_result_2ch}  that
the $c_1$ and $c_2$ values greatly outperforms the baseline $c_0 = 0$ value for almost all shift magnitudes
in the OC case of local changes. In the bottom panel for the OC case (III) of the global change,  the detection power improvement  is significant for $c_{2}$
as compared to the baseline $c_{0} =0.$ We feel this might be due to the new spatial sparsity where the profile means of only two channels have shifted. While our proposed thresholded PCA method is not designed specifically for the spatial sparsity,
the thresholding can actually take care of spatial sparsity to yield better detection power.
In addition, as compared to Figure \ref{fig:unknown_result}, Figure \ref{fig:unknown_result_2ch} implies
that  the detection powers when only $2$ out of $4$ components have changed are less than those when all $4$ components have changed.

\section{Conclusion and Future Work}

In this paper, we proposed a thresholded multivariate PCA for multichannel profile monitoring.
Our proposed method essentially conducts dimension reduction in two steps: We first apply multivariate PCA to reduce high dimensional multichannel profiles to a reasonable number of features under the normal operational state, and then use soft-thresholding techniques to further select informative features under the out-of-control state. We also give several suggestions on how to select tuning parameters based on asymptotic analysis.
Moreover, we used real forging process dataset and B-splines to build generative methods for multichannel profiles under the in-control state and
$2 \times 3 \times 7 = 42$ different out-of-control states.
Our numerical studies demonstrate that the soft-thresholding technique can significantly increase the detection power as compared to the baseline value $c_0 = 0$.

There are a number of interesting problems that have not been addressed here. From the theoretical point of view, it will be useful to investigate the efficiency of our proposed methods, and to find an optimal  value of soft-thresholding parameter $c$ that can adaptively adjust for different out-of-control states. Another direction is to investigate how to extend our proposed method to Phase II online profile  monitoring.
That will be more challenging, partly because it is more difficult to select informative principal components due to fewer out-of-control profiles since one observes profiles one at a time. Therefore, our research should be interpreted as a starting point for further investigation.

\bibliographystyle{apalike}

\bibliography{referencefile}

\newpage

\newpage

\clearpage

\clearpage

\begin{table}
\centering \caption{The value of $d_{0}$ and soft-thresholding parameters $c$'s}
\label{tab:val_c_unknown} %
\begin{tabular}{|c|c||c|c||c|}
\hline
 &  $c_{0}$  & $d_{0}$  & $c_{1}$  & $c_{2}$ \tabularnewline
\hline
OC-case (I) & 0 & 15    & 4.9  & 11.6 \tabularnewline
\hline
OC-case (II) & 0   & 9   & 7.0  & 11.6 \tabularnewline
\hline
OC-case (III) & 0 & 12    & 4.5  & 11.6 \tabularnewline
\hline
\end{tabular}
\end{table}

{\bf Note:} All our proposed methods were developed under the assumption that all $4$ components/channels are affected, 
and the magnitudes of the changes are unknown.  Table \ref{tab:val_c_unknown} lists the specific  values of $c_{0},c_{1},c_{2}$ used in our study. Note that the value of $c_0=0$ and $c_2$ do not depend on the location of the change, but the value of $c_1$ depends on the location of the change.

Note that when computing the $c_1$ value in Proposition 1, we need to know the value of $d_0,$ the number of affected principal components that are relevant to the change among a total of $d=45$ principal components. Here the value $d_0$ is chosen by the following data-driven method: We first obtain $U_{\ell,k}^{H_{0}}\{k=1,...,d\}$'s
under $H_{0}$ using the simulated in-control profiles and record the
value $A$ as the top 10\% value of $U_{\ell,k}^{H_{0}}$'s. Then, we compute
$U_{\ell,k}^{H_{1}}\{k=1,...,d\}$'s under $H_{1}$ using simulated
out-of-control profiles, and count how many $U_{\ell,k}^{H_{1}}$'s
are greater than such threshold $A$. This  gives an estimate of $d_{0}$ since it indicates the
number of altered  $U_{\ell,k}$'s if a specific fault occurs.
For the purpose of easy computation and comparison, the out-of-control scenario was conducted when all $4$ components of affected $\bftheta_{i}$ are changed, and the same $d_0$ and $c_1$ values were used in the scenario when only $2$ out $4$ components are changed.


\newpage

\newpage

\begin{figure}
 \centering

   \includegraphics[width=4in, height=1.8in]{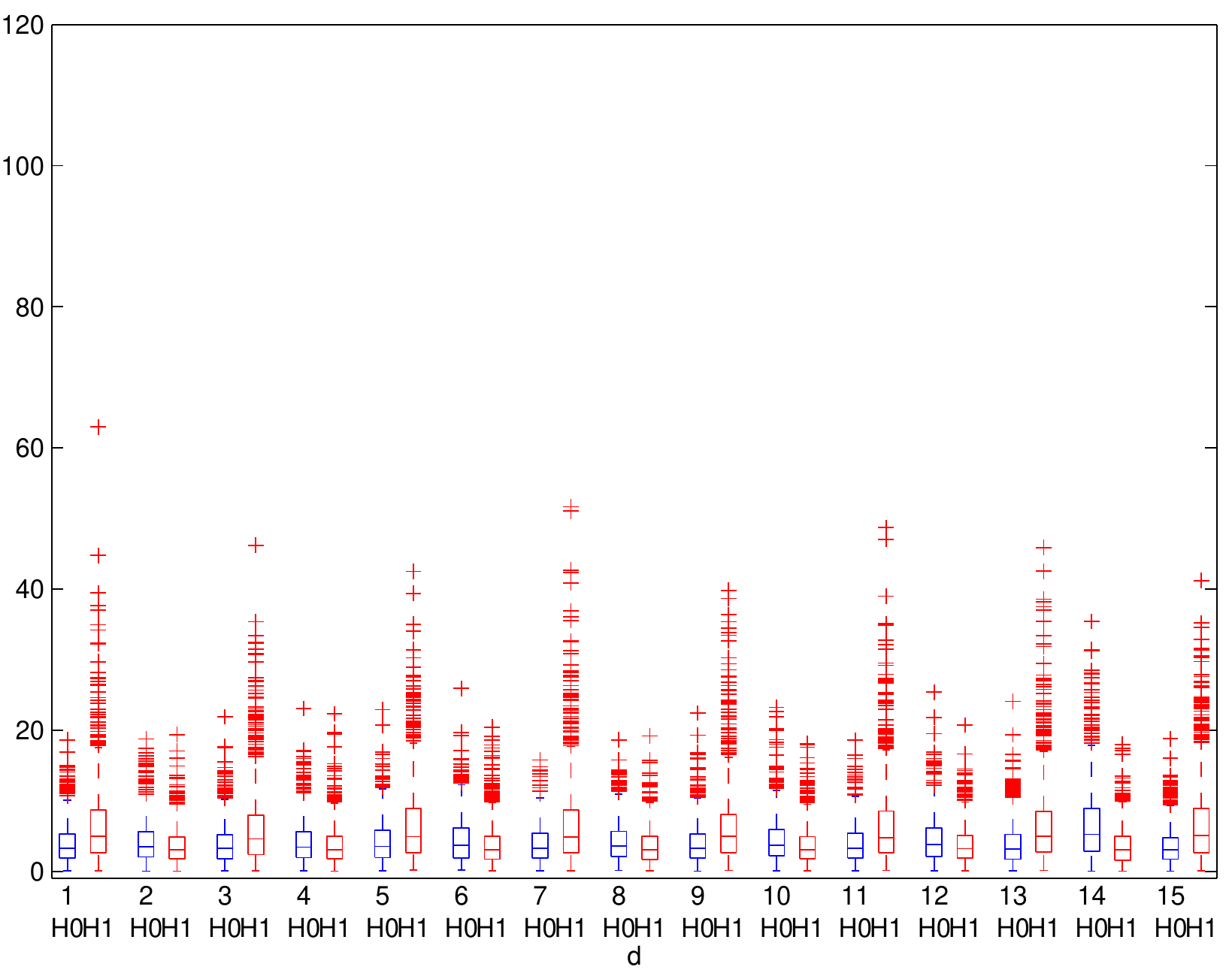}\\
  \includegraphics[width=4in, height=1.8in]{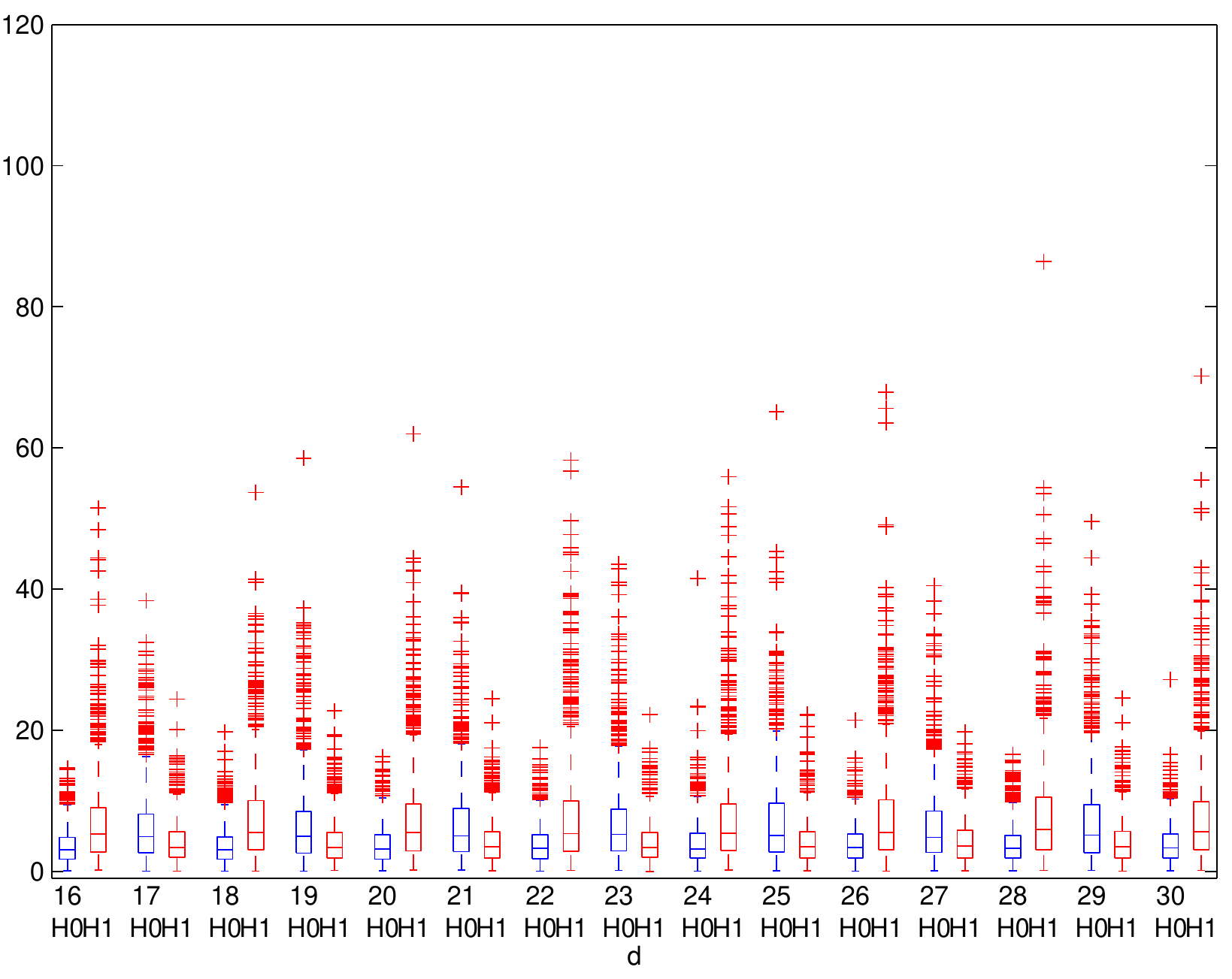} \\
  \includegraphics[width=4in, height=1.8in]{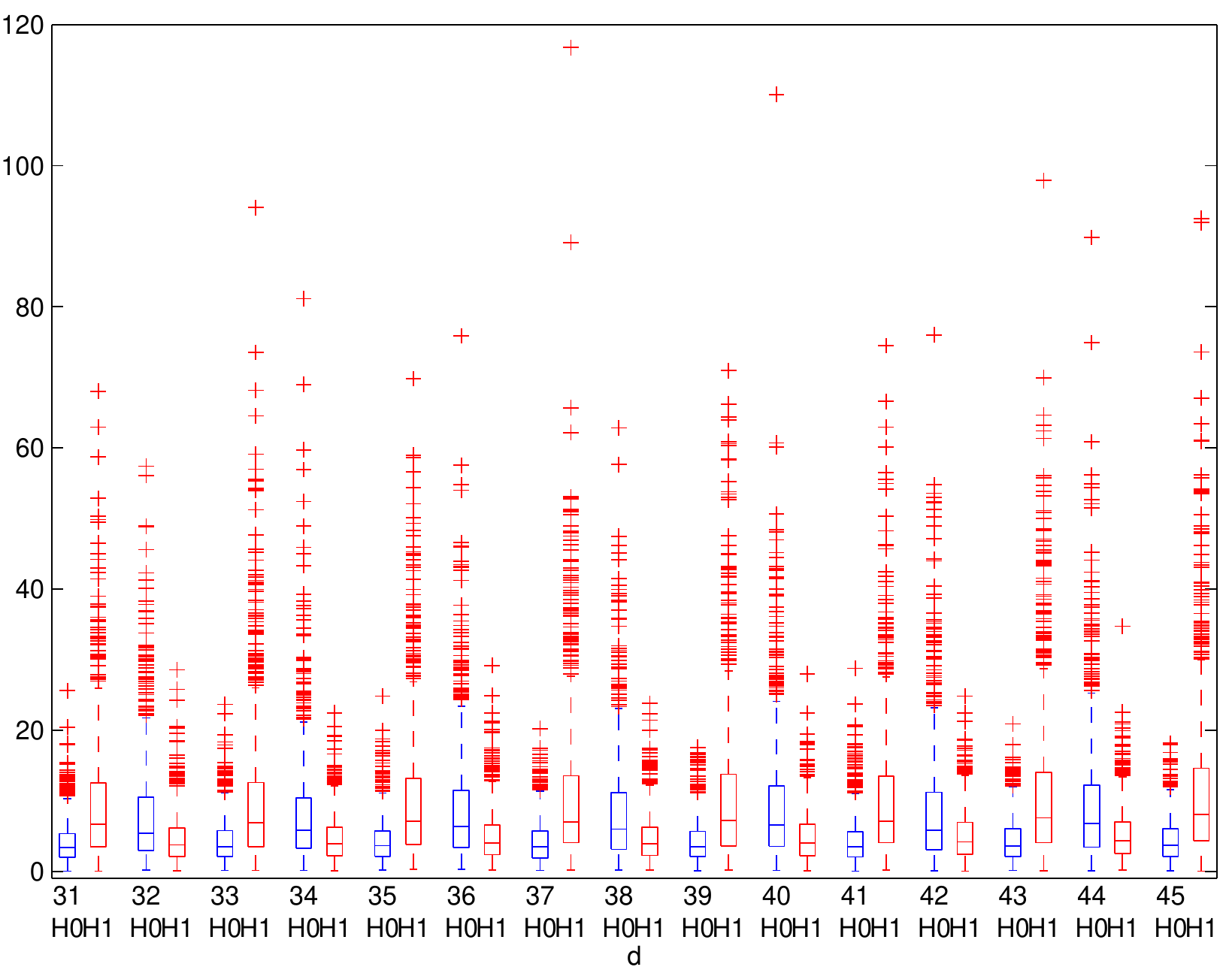}

  \caption{Box plots of  $U_{\ell=\tau=100, k}$ under the  $H_0$ hypothesis and $H_1$ hypothesis  for case (III) under all 4 channels affected scenario with $h=4$ based on 1000 replications. X axis with $k=1,...,45$ represents the projection on the $k$'th principal components. This plot implies that even for the global change, the OC distribution of the $U_{\ell,k}$'s is not necessarily stochastically larger than those IC distribution over all $k=1, \cdots, 45$ principal components. We feel that this is the reason why soft-thresholding can improve the detection power in the global change case, as it can filter out those $U_{\ell,k}$'s that have smaller OC values.
}
\label{fig:U_{lk}}
\end{figure}

\newpage
\begin{table}[]
\centering
\caption{Comparison of detection biases for each algorithms under 3 different out-of-control cases for all 4 channels affected scenario.}
\label{tab:result}
\begin{tabular}{|c|c|c|c|c|c|c|c|c|c|c|}
\hline
            &       & \multicolumn{3}{c|}{$\E(|\hat{\tau} - \tau |)$}                  & \multicolumn{3}{c|}{$\Prob(|\tau - \hat{\tau}|\le 1)$} & \multicolumn{3}{c|}{$\Prob(|\tau - \hat{\tau}|\le 3)$} \\ \hline
            & $h$ & $c_0$     & $c_1$ &       $c_2$ &       $c_0$             & $c_1$             & $c_2$            & $c_0$             & $c_1$             & $c_2$            \\ \hline
Case   & 1     & 5.18   $\pm$ 1.71 & 1.14   $\pm$ 1.89 & 0.86   $\pm$  2.09 & 0.18              & 0.15              & 0.19             & 0.40              & 0.44              & 0.36             \\ \cline{2-11}
 (I)           & 2     & 1.57   $\pm$  1.33 & 1.89   $\pm$  1.35 & 2.65   $\pm$  1.73 & 0.22              & 0.22              & 0.22             & 0.51              & 0.50              & 0.39             \\ \cline{2-11}
            & 3     & 0.95   $\pm$  1.21 & 1.51   $\pm$  1.27 & 0.59   $\pm$  1.38 & 0.27              & 0.26              & 0.25             & 0.54              & 0.54              & 0.47             \\ \cline{2-11}
            & 4     & 0.81   $\pm$  1.10 & 1.03   $\pm$  1.02 & 0.59   $\pm$  1.26 & 0.31              & 0.36              & 0.33             & 0.57              & 0.63              & 0.54             \\\cline{2-11}
            & 5     & 0.28   $\pm$  0.86 & 0.30   $\pm$  0.88 & 0.09   $\pm$  0.77 & 0.38              & 0.36              & 0.42             & 0.63              & 0.63              & 0.63             \\ \cline{2-11}
            & 6     & 0.13   $\pm$  0.73 & 0.13   $\pm$  0.79 & 0.58   $\pm$  0.47 & 0.41              & 0.42              & 0.46             & 0.65              & 0.68              & 0.66             \\ \cline{2-11}
            & 7     & 0.14   $\pm$  0.54 & 0.63   $\pm$  0.46 & 0.29   $\pm$  0.49 & 0.47              & 0.46              & 0.49             & 0.70              & 0.72              & 0.70             \\ \hline
Case  & 1     & 2.15   $\pm$  2.37 & 0.16   $\pm$  2.45 & 2.30   $\pm$  2.30 & 0.09              & 0.22              & 0.22             & 0.24              & 0.36              & 0.36             \\ \cline{2-11}
(II)           & 2     & 1.98   $\pm$  1.64 & 0.78   $\pm$  1.57 & 0.18   $\pm$  1.50 & 0.24              & 0.35              & 0.35             & 0.48              & 0.51              & 0.53             \\ \cline{2-11}
            & 3     & 1.12   $\pm$  0.88 & 0.76   $\pm$  1.08 & 1.19   $\pm$  1.23 & 0.39              & 0.40              & 0.42             & 0.60              & 0.61              & 0.63             \\ \cline{2-11}
            & 4     & 0.11   $\pm$  0.70 & 0.67   $\pm$  0.78 & 0.43   $\pm$  0.71 & 0.48              & 0.53              & 0.56             & 0.72              & 0.76              & 0.76             \\ \cline{2-11}
            & 5     & 0.51   $\pm$  0.62 & 0.02   $\pm$  0.54 & 0.24   $\pm$ 0.57 & 0.58              & 0.67              & 0.65             & 0.78              & 0.83              & 0.86             \\ \cline{2-11}
            & 6     & 0.22   $\pm$  0.53 & 0.51   $\pm$  0.49 & 0.49   $\pm$  0.49 & 0.70              & 0.76              & 0.73             & 0.86              & 0.91              & 0.90             \\ \cline{2-11}
            & 7     & 0.50   $\pm$  0.48 & 0.02   $\pm$  0.14 & 0.07   $\pm$  0.16 & 0.77              & 0.81              & 0.80             & 0.90              & 0.95              & 0.95             \\ \hline
Case & 1     & 0.07   $\pm$  1.13 & 0.16   $\pm$  1.07 & 1.18   $\pm$ 1.25 & 0.35              & 0.34              & 0.31             & 0.57              & 0.57              & 0.50             \\  \cline{2-11}
  (III)     & 2     & 0.58   $\pm$  1.11 & 0.37   $\pm$  1.01 & 0.52   $\pm$  1.07 & 0.39              & 0.35              & 0.34             & 0.60              & 0.61              & 0.54             \\  \cline{2-11}
            & 3     & 0.85   $\pm$  1.05 & 0.67   $\pm$  0.94 & 0.51   $\pm$  0.84 & 0.43              & 0.39              & 0.38             & 0.64              & 0.61              & 0.56             \\  \cline{2-11}
            & 4     & 0.15   $\pm$  0.90 & 0.11   $\pm$  0.73 & 0.43   $\pm$  0.82 & 0.45              & 0.41              & 0.40             & 0.67              & 0.64              & 0.61             \\  \cline{2-11}
            & 5     & 0.13   $\pm$  0.84 & 0.11   $\pm$  0.66 & 0.27   $\pm$  0.55 & 0.47              & 0.45              & 0.45             & 0.69              & 0.68              & 0.66             \\  \cline{2-11}
            & 6     & 0.44  $\pm$ 0.79 & 0.04   $\pm$  0.58 & 0.03   $\pm$  0.52 & 0.49              & 0.48              & 0.46             & 0.70              & 0.71              & 0.67             \\  \cline{2-11}
            & 7     & 0.39   $\pm$  0.63 & 0.15   $\pm$  0.54 & 0.01   $\pm$  0.53 & 0.51              & 0.49              & 0.47             & 0.72              & 0.72              & 0.67             \\ \hline
\end{tabular}
\end{table}

\newpage

\end{document}